Article title: Smart Spectral Doppler Ultrasound Imaging using Motion Tracking based on Wide-beam transmission and Neural Networks: Preliminary Results

Authors: Jihye Baek[1], Allison Lee[2], Jinbum Kang[3], Yangmo Yoo[3]

[1]Department of Radiology, Stanford University, Palo Alto, CA, 94304, USA

[2]Department of Electrical Engineering, Stanford University, Palo Alto, CA, 94305, USA

[3]Department of Electronic Engineering, Sogang University, Seoul 04107, Korea

Running title: 2D motion tracking doppler

**Abstract**

Ultrasound spectral Doppler imaging (USDI) provides quantitative blood flow information for vascular diagnosis, but its accuracy is limited by the one-dimensional (1D) sample volume (SV). Motion from respiration, patient movement, and probe manipulation can move the target vessel out of the SV, causing spectral broadening, signal reduction, or loss of the spectrum. A two-dimensional (2D) SV can address this, but sequential focused-beam transmission reduces the pulse-repetition frequency (PRF), and plane-wave transmission lowers the signal-to-noise ratio (SNR). In this study, we propose a wide-beam transmission method that acquires a 2D SV from a single transmit event while preserving the PRF. A Hamming network, which requires no prior training, tracks axial and lateral motion within the 2D SV and updates the position of an adaptive 1D SV for each frame. Field II simulations showed that the wide-beam transmission had a spectral SNR comparable to that of focused-beam transmission and approximately 10 dB higher than that of plane-wave transmission. The motion tracking correctly detected the target position in over 93% of frames for vessel radii greater than 1 mm, and in 100% of frames for radii greater than 2 mm. *In vivo* studies of a carotid artery (linear array) and a hepatic vein (convex array) showed that the proposed method recovered obscured spectra and increased the spectral SNR by 9.3 dB and 4.5 dB, respectively. These results demonstrate that the proposed wide-beam transmission with Hamming network–based tracking can improve the accuracy of spectral Doppler imaging under tissue motion. Moreover, we anticipate that the proposed approach can be implemented in real-time due to its single transmission and simple two-layer architecture with low computational complexity.

## I. INTRODUCTION

Ultrasound spectral Doppler imaging (USDI) has been widely used for diagnosing various vascular diseases as it provides quantitative analysis of blood flow, including velocity profile and functional indices (e.g., resistive index, RI). This information plays a significant role in evaluating atherosclerosis in carotid and renal arteries [1, 2] and in examining fetal hemodynamics [3]. However, the clinical usefulness of USDI tends to depend on the quality of the displayed image (i.e., velocity profile as spectrogram) since both the spectral Doppler examination and the subsequent diagnostic decision are performed based on the observation of the spectrogram. Thus, improving USDI accuracy is essential to enhancing its clinical usefulness.

Several factors affect USDI accuracy. Current USDI uses a one-dimensional (1D) range gate, called a sample volume (SV), to acquire ultrasound signals backscattered by blood scatterers. Nonstationary movement caused by respiration, patient movement, and probe manipulation by operators (e.g., sonographer) can move the target blood flow outside the 1D SV. This can cause spectral broadening, signal reduction, or complete loss of the spectrum. Moreover, lateral motion renders obtaining blood flow with 1D SV impossible.

A 2D SV could enable tracking in both axial and lateral directions during the spectral Doppler examinations. A 2D SV for USDI can be obtained by sequentially transmitting multiple focused beams to cover multiple scanlines, around the target vessel, as in ultrasound color Doppler imaging. However, sequential transmission of multiple focused beams substantially reduces the effective pulse-repetition frequency (PRF), which limits the maximum measurable velocity. In addition, time delays between transmissions can cause decorrelation among scanlines [4].

Recently proposed uncompounded plane-wave transmission techniques can be used to obtain a 2D SV in USDI [5-7]. Uncompounded plane-wave transmission avoids the PRF reduction and decorrelation, but it has a lower signal-to-noise ratio (SNR) than focused transmission. Coherently compounding multiple tilted plane-waves can improve SNR, but it reduces the effective PRF and can introduce decorrelation during coherent compounding. Alternatively, a 2D SV can be obtained using weakly-focused wave transmission, in which a less focused wave with a larger F-number is transmitted, and multiple scanlines, such as four or six, are simultaneously reconstructed using parallel beamforming techniques [8]. This approach offers advantages over a sequential focused-beam transmission in terms of PRF and decorrelation. However, using fewer transmit elements to obtain a larger F-number reduces the transmitted power and can lower SNR [9].

Effective motion tracking is important for USDI with the incorporation of 2D SV. Speckle tracking methods based on cross-correlation or sum-absolute-difference between adjacent scanlines have been traditionally used in ultrasound B-mode imaging [10]. However, these methods can be less suitable for tracking vessel motion because small vessels may have vaguely defined features. Deep neural networks (DNNs), which have been used to improve image quality, assist diagnosis, and classify disease types in ultrasound B-mode and Doppler imaging [11-13], can be applied to trace vascular motion in USDI with a 2D SV. However, their need for large training datasets may limit their suitability for this application. A Hamming network [14, 15] requires no prior training or large training dataset. In addition, the Hamming network has simple architecture that makes real-time implementation on modern ultrasound machines possible.

In this paper, we propose a wide-beam transmission method for USDI with a 2D SV. A broad beam is transmitted with a small F-number, similar to a focused-beam transmission, and multiple scanlines corresponding to 2D SV in USDI are reconstructed simultaneously at PRF intervals. Hamming network uses these multiple scanline data for tracking vascular motion during Doppler examination. Section II.A and II.B introduces the proposed wide-beam transmission and motion tracking based on the Hamming network, respectively. Section II.C describes experimental setup, and Section II.D defines the evaluation metrics. Section III.A and III.B present the Field II simulation and *in vivo* study, respectively. Section IV discusses the findings, and Section V presents the conclusion.

## II. METHODS

**Figure 1** compares conventional USDI with the proposed 2D SV-based USDI. Conventional USDI transmits a focused-beam toward a 1D SV and uses the received signal for pulsed-wave Doppler processing, i.e., short-time Fourier transform. The proposed method transmits a wide-beam toward a 2D SV. Then, within the 2D SV, motion tracking is performed to determine the local position of an adaptive 1D SV. The backscattered signal from the adaptive 1D SV is then processed using the same Doppler processing steps as in conventional USDI.

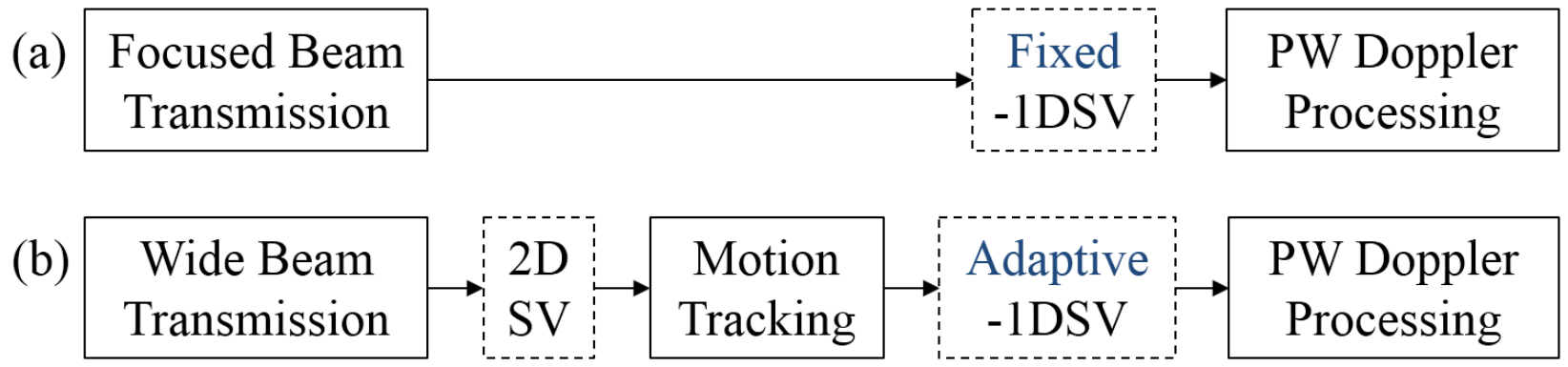

**Figure 1.** Schematic diagrams of (a) the conventional spectral Doppler processing and (b) the proposed 2D motion tracking Doppler processing.

### A. Wide-beam transmission

We propose a wide-beam transmission method. To describe its transmition, we define a single delay equation that unifies three different cases: the conventional focused beam, the proposed wide-beam, and the conventional plane wave. **Figure 2** illustrates these three transmission configurations.

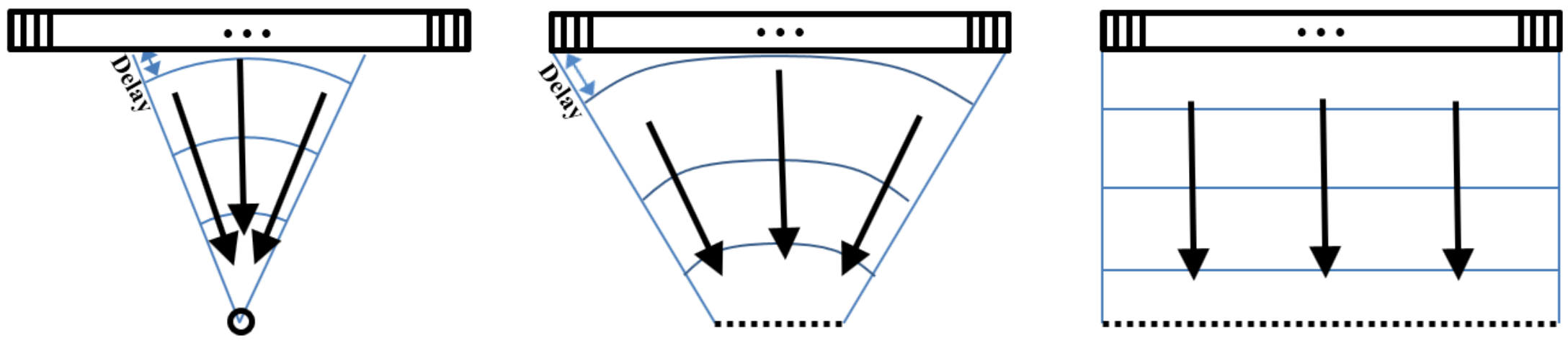

**Figure 2. Three configurations for a single transmit event: focused beam, proposed new wide-beam, and plane wave, from left to right.**

We first define $L_f$ as the length of the focal line at the focal depth. For the focused-beam transmission, $L_f = 0$ because the beam is assumed to converge to a single focal point. For plane-wave transmission, $L_f$ equals transducer width D. Lastly, for the wide-beam, $L_f$ equals the beamwidth that can be determined by users between 0 and D. The specific cases of $L_f = 0$ and $L_f = D$ correspond to focused-beam and plane-wave transmission, respectively.

Focused-beam transmission uses a subset of transducer elements determined by the F-number and focal depth. We define the active aperture size A as the lateral span of the active elements. **Table 1** summarizes the parameters ($L_f$, D, and A) for the three transmission methods. The transmit-delay expression $\Delta t_d$ is given by:

$$\Delta t_d = \sqrt{\left(x_e - x_f\right)^2 + {z_f}^2} - \min\left(\sqrt{\left(x_e - x_f\right)^2 + {z_f}^2}\right) \quad (1)$$

where the position of each transducer element in the transducer is $(x_e, 0)$:

$$x_e \in \left\{x_e \middle| e = e_i < e_{i+1} = e_i + 1 < \cdots < e_{i+s-1} = e_i + s - 1 \in \text{Integer},\ x_{e_{i+s-1}} - x_{e_i} = A,\ x_e - x_{e-1} = \frac{A}{s-1}\right\}$$

(2)

where A is the active aperture size and s is the number of active elements in the transducer. With A defined as the distance between the first and last active element centers, the element spacing of the transducer is $\frac{A}{s-1}$. E ranges from $e_i$ to $e_{i+s-1}$ where $1 \le e_i\ and\ e_{i+s-1} \le n_t$, and $n_t$ is the total number of transducer elements.

The corresponding focal-points coordinates are $\left(x_f, z_f\right)$:

$$x_f \in \left\{x_f \middle| f = e\ for\ all\ x_e,\ x_{f_{i+s-1}} - x_{f_i} = L_f, x_f - x_{f-1} = \frac{L_f}{s-1}\right\} \quad (3)$$

where $L_f$ is the length of the focal line or a point at a focal depth $z_f$, and $s$ and $e$ have the same meaning as in the definition of $x_e$. $D$ is the full aperture size. For example, focused-beam transmission with 32 activated elements has s =32 and $e = f = 49, 50, \ldots, 80$. Wide-beam transmission using all 128 elements has s =128 and $e = f = 1, 2, \ldots, 128$.

| Transmission | Active aperture size (A) | Length of a focused line at a focal depth ($\boldsymbol{L_f}$) |
|---|---|---|
| Focused beam | $A < D$ | $L_f = 0$ |
| Wide-beam | A = D | $0 < L_f < D$ |
| Plane wave | $A = D$ | $L_f = D$ |

**Table 1**. Transmission conditions for focused beam, new wide-beam, and plan-wave transmission.

In this study, we used the wide-beam transmission with a beam width $L_f$ of approximately 1-cm to cover the motions typically encountered during USDI. Measurements of abdominal respiratory motion [16] reported an average amplitude of is 2.17-mm during natural breathing and a maximum of 4.23-mm under unstable physical or emotional states (e.g. stress and vocalization). A kidney-tracking study [17] reported a motion amplitude of approximately 3-mm due to organ and respiration. These reported amplitudes suggest that a beam width of 1-cm can sufficiently cover target motion during spectral Doppler imaging, including motion in liver and kidney.

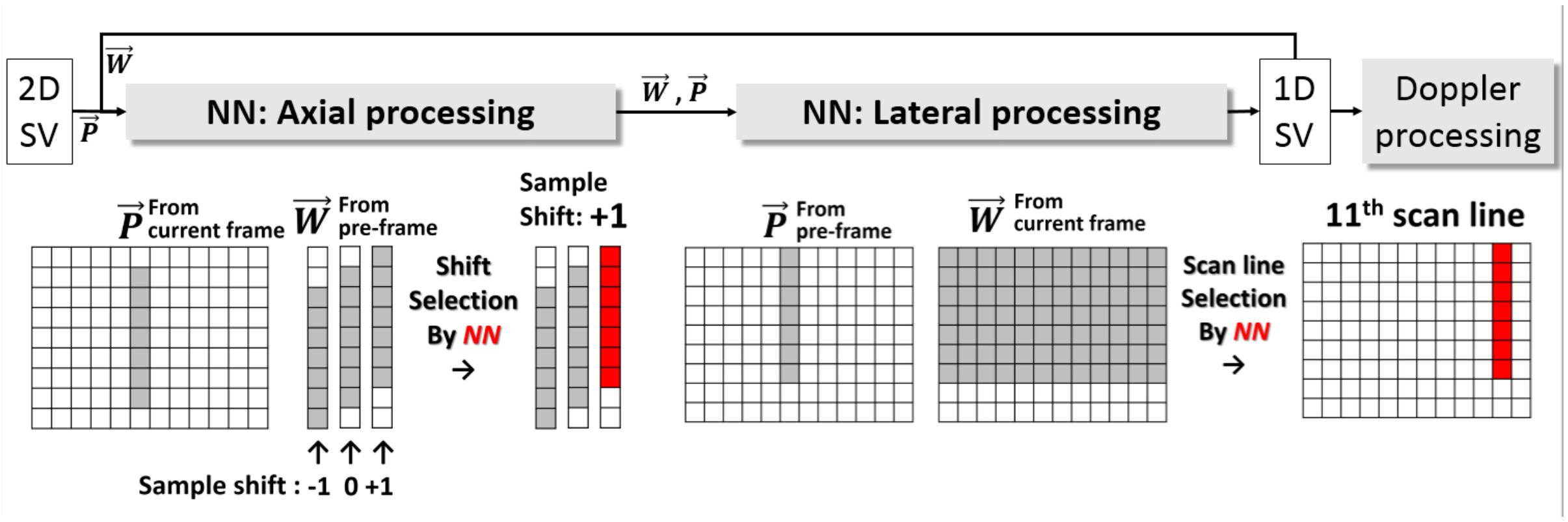

**Figure 3.** Proposed motion tracking process. Axial and lateral processing are performed sequentially within a 2D SV. Axial processing selects the sample shift that accounts for motion in depth. Lateral processing then selects the scanline, defining an adaptive-1D SV whose signal is used for Doppler processing.

### B. Motion tracking based on neural networks

Motion tracking uses ultrasound signals within a 2D SV acquired with the wide-beam transmission method described in **Section II.A**. **Figure 3** depicts the Hamming network-based motion tracking performing in the axial and lateral directions. The network tracks the flow target in both directions to determine the range gate positions in each frame; this range gate is termed the adaptive-1D SV. Conventional USDI uses a fixed-1D SV, whereas the proposed method updates the adaptive-1D SV within the 2D SV by tracking both axial and lateral directional motions.

The Hamming network locates the axial position of a flow target so that the adaptive-1D SV covers the flow in axial direction. It then locates the lateral position of the flow, and the corresponding scanline is set as the lateral position of the adaptive-1D SV. The signal within the resulting adaptive-1D SV is used to estimate conventional spectral Doppler velocity.

The Hamming network determines the axial and lateral positions of the adaptive-1D SV by calculating correlations between current and previous frames. **Figure 4** illustrates its two layers: a feedforward layer and a competitive layer. The feedforward layer computes correlation based on inner products, and the competitive layer selects a maximum value among the results of the feed forward layer.

As shown in the **Figure 3**, a weight matrix $\vec{W}$ and a prototype vector vector $\vec{p}$ are defined for both axial and lateral processing, $\vec{W}$ is a 2D matrix and $\vec{p}$ is a column vector. For the axial processing, $\vec{p}$ contains sample from the center scanline of the current frame, over the axial range of the previous frame's 1D SV. The matrix $\vec{W}$ is defined using (1) 1D SV from the previous frame and (2) axially shifted versions of that 1D SV; for example, the 1$^{st}$, 2$^{nd}$, and 3$^{rd}$ rows of $\vec{W}$ can represent shifts of -1, 0, and +1 sample of the 1D SV from the previous frame. Hamming network selects how many sample shifts from the previous to reach greatest similarity to the current frame ($\vec{p}$). Therefore, as illustrated in Figure 3 in NN: Axial processing, the shift selection by Hamming network determined a best correlated shift, which can be an index of a column within $\vec{W}$; the selected shift in axial direction is considered as movement from previous to current frame. Thus, the shifted axial position becomes the axial position of the adaptive-1D SV, which indicate the red highlighted samples in the Figure 3 in NN: Axial processing. The ultrasound signal within the selected samples and the center scanline becomes $\vec{p}$, which is acquired from the previous frame and used for lateral processing. $\vec{W}$ for the lateral processing is obtained within 2D SV from current frame. $\vec{W}$ includes all scanlines within the 2D SV; whereas it only contains the same sample range with the $\vec{p}$ from the previous frame because we already selected axial region of interest in the axial processing. The ultrasound signals within the 2D area (gray-shaded) became $\vec{W}$. Lastly, the Hamming network finds a scanline to determine adaptive-1D SV by selecting the best correlated scanline within $\vec{W}$. That way, by shifting the sample or scanline position of adaptive-1D SV, the Hamming network can select a shifted sample and scanline position, which shows the best correlation between frames at the adaptive-1D SVs. Since the target is likely to move several scanlines or samples between the continuous two frames, which can be covered within 2D SV, the flow target motion can be tracked using this approach. For example, in **Figure 3**, axial directional shift of +1 sample and 11$^{th}$ scanline in 2D SV of current frame shows the greatest correlation with the 1D SV of the previous frame, so that the axial movement of the target is +1 sample, and the lateral location of the target is changed from center scanline to 11th scanline.

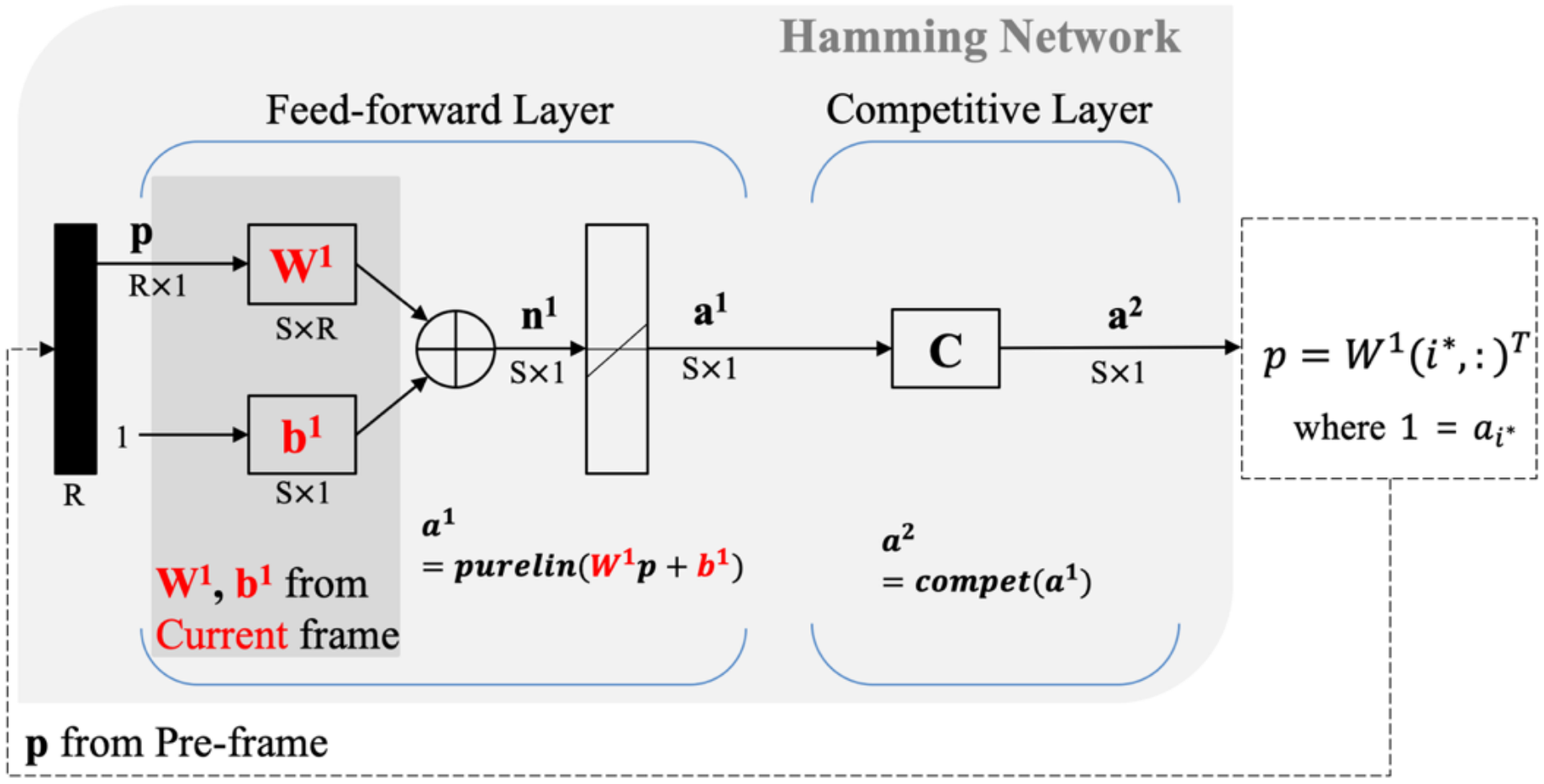

**Figure 4**. Hamming network for motion tracking. The network updates the position of the adaptive-1D SV in each frame to follow motion during ultrasound scanning.

To construct the components of prototype ($\vec{p}$) and weight ($\vec{W}$) matrices, we use tissue signals, blood flow, and their derivatives as inputs of the Hamming network. The absolute values of complex demodulated (IQ) signal represent tissue signals, and the clutter-filtered data (CL) represent blood flow. In addition, their partial derivatives with respect to the axial direction can provide more information about whether the IQ and CL increase or decrease, which can add more details compared to using only the magnitudes of IQ or CL. These four types of features generate a multiple-input neuron $\vec{S}$ for an input of the Hamming network:

$$\vec{S} = \left[s_1\ s_2\ \cdots\ s_{4sp-2}\right]^T = \left[IQ_1\ \cdots\ IQ_{sp}\ CL_1\ \cdots\ CL_{sp}\ \frac{\partial IQ}{\partial x}_1 \cdots \frac{\partial IQ}{\partial x}_{sp-1}\ \frac{\partial CL}{\partial x}_1 \cdots \frac{\partial CL}{\partial x}_{sp-1}\right]^T \quad (4)$$

where $sp$ is the number of samples in the 2D SV and $x$ denotes the axial direction.

In contrast to general neural networks that require prior training to create weight matrices, the weight matrix $\vec{W}$ of Hamming network can be set by given several prototype patterns $\vec{G}$; the prototype patterns are determined by the four types of information as introduced in eq (4). The weight matrix $\vec{W}$ and prototype patterns $\vec{G}$ are as follows:

$$\vec{W} = \begin{bmatrix} \vec{P}_1^T \\ \vec{P}_2^T \\ \vdots \\ \vec{P}_Q^T \end{bmatrix}, \vec{G} = \{\vec{P}_1, \vec{P}_2, \cdots, \vec{P}_i, \cdots, \vec{P}_Q\} \quad (5)$$

where Q denotes the total number of scan lines or sample shifts. $\vec{P}_i$ is determined according to eq (4) by using i$^{th}$ scan line or i$^{th}$ sample shift data. Therefore, it doesn't require prior training to determine weight vectors. When the Hamming network determines where to move laterally the 1D SV from the previous frame to the current frame among M scanlines, the weight matrix has M prototype patterns: Q = M. Similarly, in the axial processing, when the network determines sample shift among –M, -(M-1), …, -1, 0, 1, …, M samples, the weight matrix has 2M+1 prototype patterns: Q = 2M+1.

By using these input and weight matrices, the outputs of the feed forward layer are the inner products between the input vector and each prototype vector of the weight matrix; the number of output elements is equal to the number of prototype patterns.

Feedforward layer:

$$\vec{W} \cdot \vec{S} + \vec{b} = \begin{bmatrix} \vec{P}_1^T \vec{S} \\ \vec{P}_2^T \vec{S} \\ \vdots \\ \vec{P}_Q^T \vec{S} \end{bmatrix} + \begin{bmatrix} \sum CL_1 \\ \sum CL_2 \\ \vdots \\ \sum CL_Q \end{bmatrix} \quad (6)$$

where we define $\vec{b}$ is a bias and a sum of clutter filtered data for each prototype, meaning that the greater the blood flow, the larger the value. It enables the method to select a region of sufficient flow, which should become an input for USDI.

The output of the feed forward layer is used as an input of the competitive layer. The original competitive layer is called the recurrent layer since the procedure is iterative, selecting a single index of prototype patterns. To reduce processing time and computational complexity in recurrent processing, we modified the competitive layer so that it selects the index of maximum value among the input vector elements of the competitive layer. The greater the correlation of input with prototype vectors, the larger the value of the inner product, whereby the Hamming network can select which prototype vector is the most correlative with input vector. Although there is the trade-off between the accuracy and speed for the selection, simply choosing maximum value instead of using the recurrent layer is enough to determine the movements correctly based on the ultrasound signal. Therefore, the Hamming network determines the position of 1D SV within the 2D SV for every frame, and then the 1D SV can be used as an input of Doppler processing.

**C. Experimental setup**

The proposed method was evaluated using Field II simulations [18] and *in vivo* studies, and the results were compared with those of conventional USDI. Raw RF channel data were acquired using Field II and an ultrasound scanner (Vantage 128, Verasonics Inc., Kirkland, WA, USA). A 5.2-MHz linear array transducer and a 3.1-MHz convex array transducer (Verasonics Inc., Kirkland, WA, USA) were used for the *in vivo* studies. Virtual transducers with the same specifications were used in Field II simulations. The proposed transmission method described in Section II.A was implemented in both studies. The neural network-based motion tracking approach described in Section II.B was implemented in MATLAB (Mathworks, Natick, MA, USA).

In the Field II simulation study, we modeled a virtual vessel undergoing global motion. The vessel radius was 1.5-mm, and the wall thickness was 10% of the radius. Flow within the vessel was parabolic and laminar, as shown in the **Figure 5 (a)**. The flow RI was set to 0.6, based on normal renal [2] and the carotid arterial [1] RI values. Vessel movements was simulated as the global motion based on human respiration patterns [16] and detected kidney motion [17], with a motion frequency of 0.22-Hz.

The performance of the proposed method was evaluated separately for the axial and lateral processing. For the axial simulation, the beam-to-flow angle and a peak velocity were 60 degrees and 35 cm/s, respectively. For the lateral simulation, these values were 30 degrees and 20-cm/s, respectively. As a result, the velocity components along the beam direction were approximately equal, at 17-cm/s, for both simulations. **Figures 5(b) and 5(c)** show the simulated global motion in the axial and lateral directions, respectively.

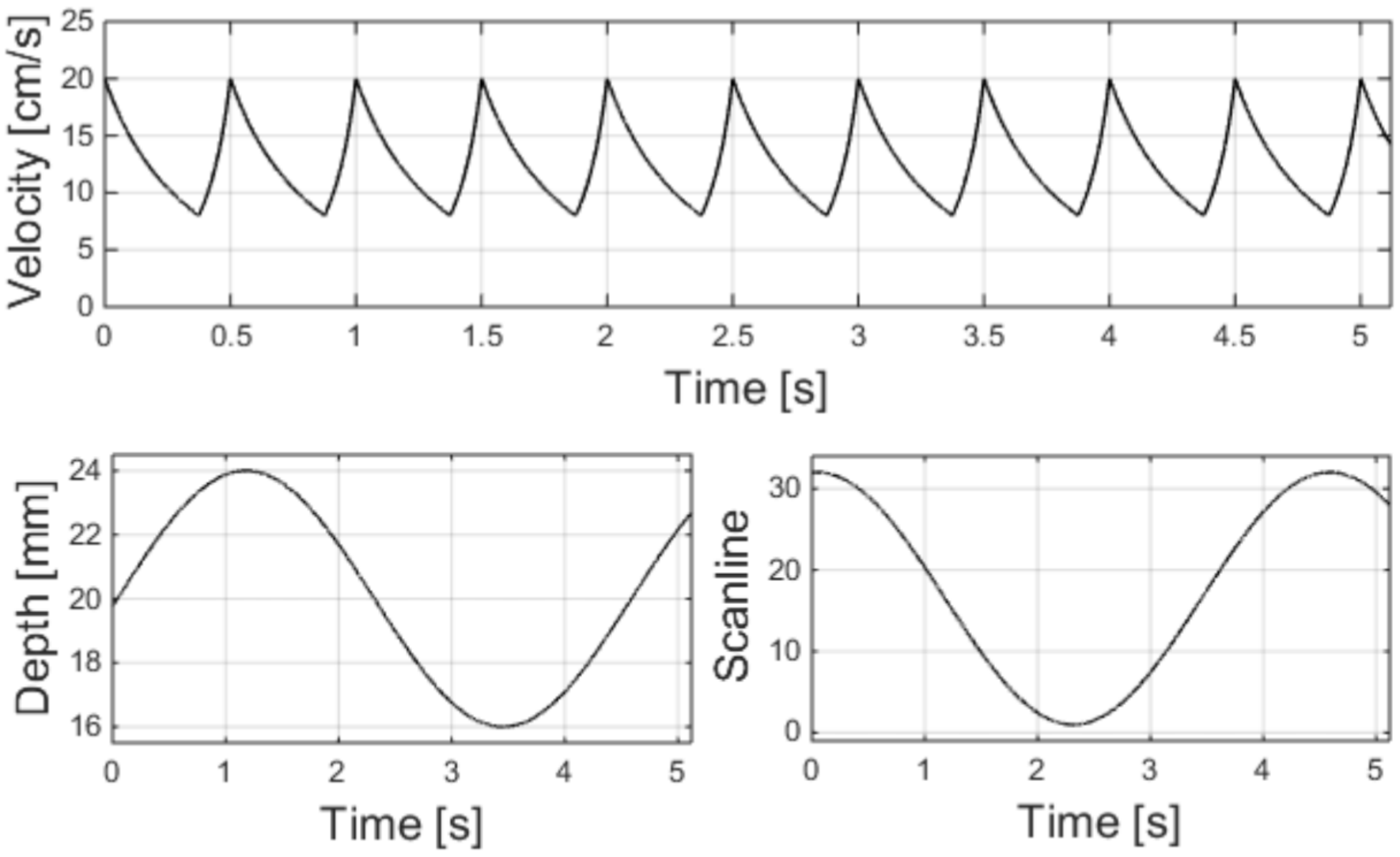

**Figure 5**. Simulation inputs: (a) pulsatile blood flow, (b) axial global motion, and (c) lateral global motion.

Healthy volunteers were enrolled in the *in vivo* feasibility study, and their Doppler spectra were obtained from 2D SVs acquired using wide-beam transmission. A carotid artery and liver hepatic veins were scanned using the linear and convex array transducers, respectively. Following wide-beam transmission, 2D SVs were reconstructed with axial and lateral dimensions of approximately 1-cm.

For the carotid artery, the 2D SV extended from a depth of 1.3-cm to 2.3-cm depth and had a lateral extent of 0.95-cm, with 32 scanlines. For the hepatic veins, the 2D SV extended from depth of 4.5-cm to 5.5-cm and had a lateral extent of 10.7-mm, with 12 scanlines. These 2D SVs were used for motion tracking and velocity estimation.

**D. Evaluation Metrics**

Performance was assed qualitatively and quantitatively. For qualitative assessment, we compared the shape of spectra obtained using the conventional and proposed method. Quantitative assessment used three metrics: (1) differences between simulated and measured positions, (2) neural network classification accuracy, and (3) spectral SNR.

First, the accuracy of the axial and lateral processing was evaluated by calculating position differences in the depth and scan line direction, respectively:

$$\text{Depth Difference [mm]} = |(Simulated\ axial\ position) - (NN\ result\ of\ axial\ processing)|$$

$$\text{Scanline Difference [scanline]} = |(Simulated\ lateral\ position) - (NN\ result\ of\ lateral\ processing)|$$

where the simulated axial and lateral positions were determined by the global motion described in Section II.C. These differences can indicate the accuracy of the neural network method, with smaller values indicating greater accuracy.

Second, we calculated the percentage (%) of frames in which the SV selected by the neural network was positioned within the simulated vessel during one period of simulated motion. A frame was classified correct if the selected SV was within the vessel and incorrect if it was outside the vessel. The percentage of correct classifications was calculated as follows:

$$\%\ \text{Correct classification} = 100 \times \frac{Number\ of\ frames\ that\ is\ correctly\ classified}{Number\ of\ total\ frames} \tag{7}$$

where the total number of frames is the sum of frames classified as correct and incorrect.

Finally, spectral SNR was calculated from the PW Doppler images as follows:

$$\text{SNR}(t) = 10 \times \log_{10}\left(\frac{P_{signal}(t)}{P_{noise}(t)}\right) \tag{8}$$

where $P_{signal}(t)$ and $P_{noise}(t)$ are the powers of the spectral signal and noise in spectral Doppler imaging, respectively. $P_{signal}(t)$ is defined as the maximum power at the time $t$, and $P_{noise}(t)$ is estimated by averaging the noise signals at the time $t$.

## III. RESULTS

### A. Field II simulation

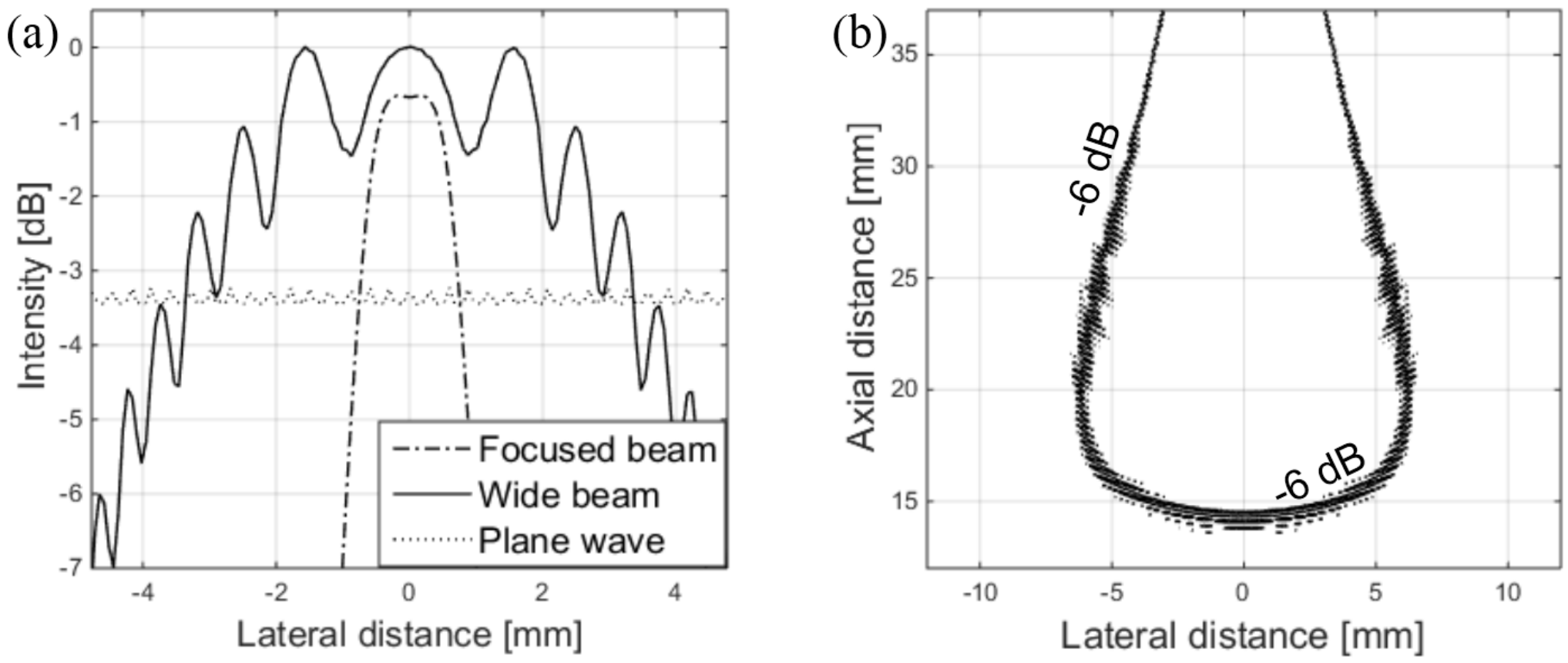

**Figure 6**. (a) Lateral beam profiles for focused-beam, wide-beam, and plane-wave transmission. (b) The -6 dB contour relative to the maximum beam intensity of wide-beam.

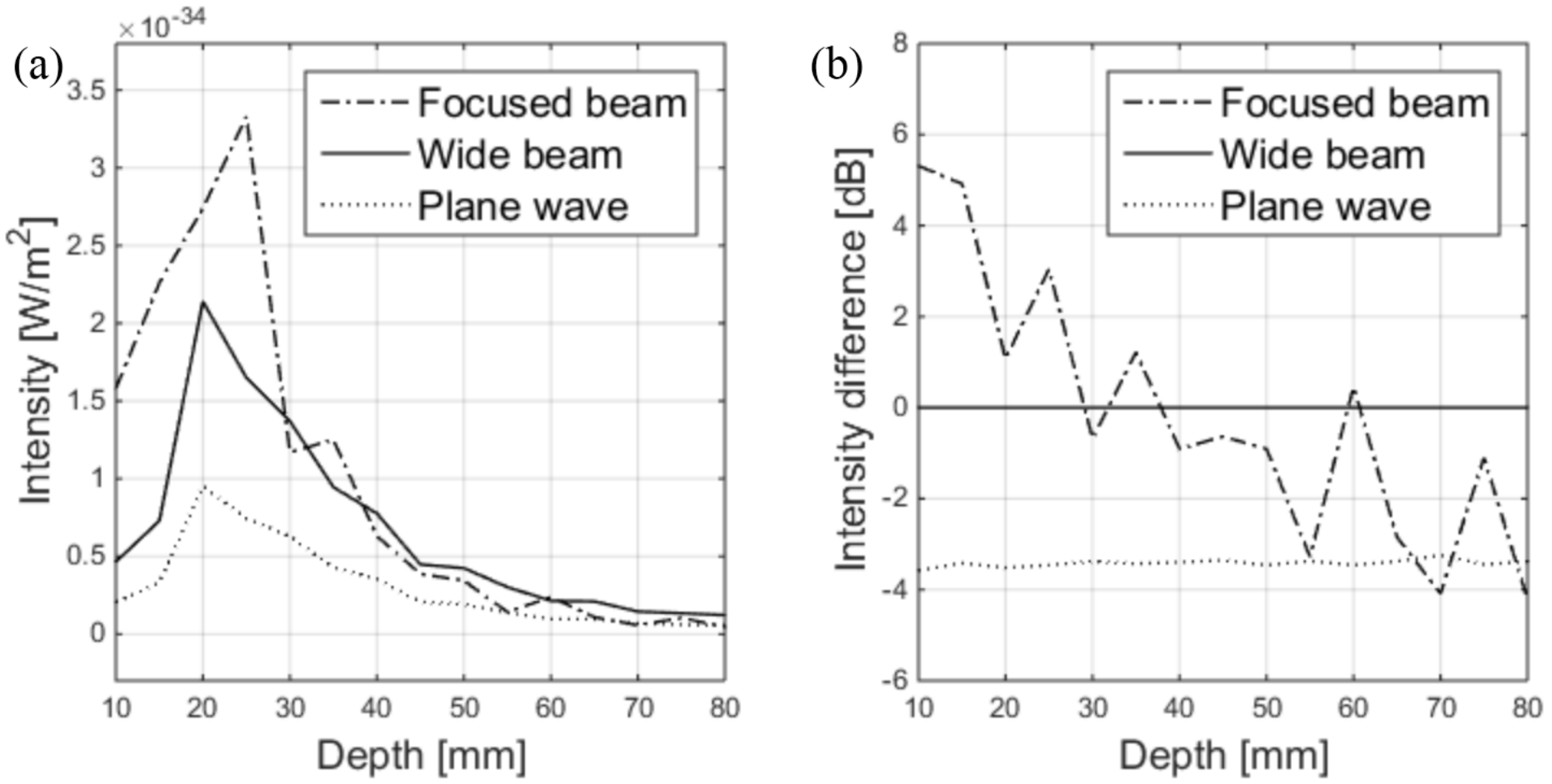

**Figure 7**. Beam energy comparison of focused-beam, wide-beam, and plane-wave transmission. (a) Intensity as a function of depth. (b) Relative intensity difference in dB, using the wide-beam intensity at each depth as the 0 dB reference.

The proposed new wide-beam method was simulated in Field II, and the results are shown in **Figures 6 and 7**. 5.2 MHz center frequency ultrasound was transmitted. As described in **Section 2. A.**, the active aperture size A is equal to the full aperture size including 128 elements of the transducer, and the length of a focused line $L_f$ at a focal depth of 30 mm is 9.5 mm including 32 elements.

**Figure 6 (a)** shows the beam profiles along with lateral distance for the three methods: focused beam, new wide-beam, and plane wave. For the wide-beam, the relative intensity for 1 cm of lateral length is within -6 dB range when compared with the maximum intensity of the focused beam, and at the center of the beam, the wide-beam has greater intensity than focused beam; meanwhile, it has higher energy than the plane wave for approximately 0.75 cm range. **Figure 6 (b)** illustrates the -6dB contour from the maximum intensity of the wide-beam, and thus the inner region can offer valid ultrasound intensity to estimate Doppler velocity. The area within the -6 dB contour covers wider lateral area compared to that of conventional focused beam. **Figure 7 (a)** represents peak energy along with depth for the three methods. Similar to the lateral distance, the intensity of the wide-beam is comparable to that of focused beam; on the other hand, the energy is higher than that of the plane wave. **Figure 7 (b)** shows relative dB scale intensity, which is obtained by the maximum values of the wide-beam method at each depth to check the relative intensities when compared with the wide-beam: the dB scale difference form that of the wide-beam. The difference between the focused and the wide-beam is under 5.3 dB with the absolute average of 2.5dB. The intensity of the plane wave tends not to exceed that of the wide-beam, and the average value difference between them is 3.4 dB. Therefore, the wide-beam transmission can deliver higher and comparable intensity when compared with plane wave and focused beam, respectively. **Figure 8** depicts SNR of spectra of the three methods. SNR of the wide-beam approach is comparable to that of focused beam and approximately 10 dB higher than that of plane wave.

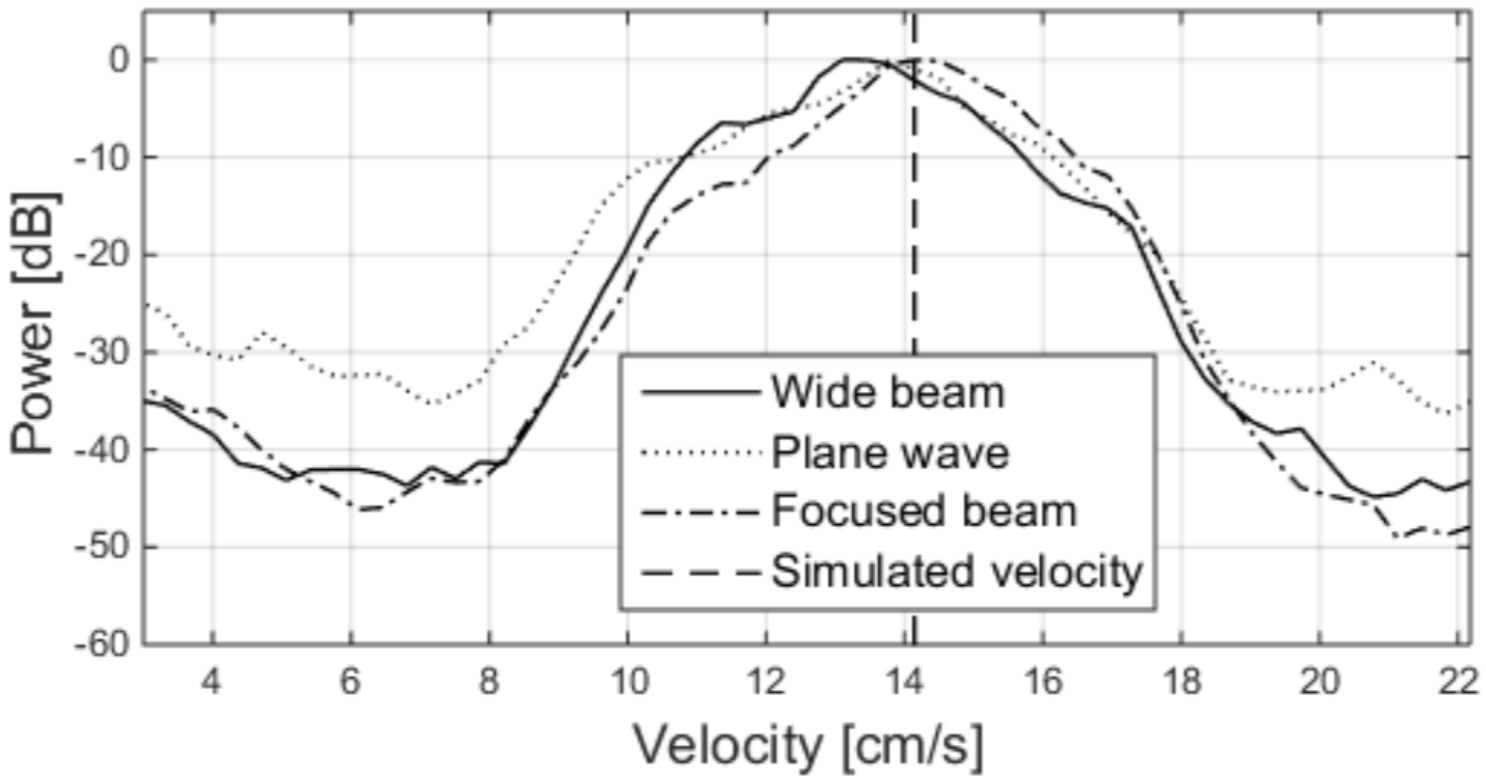

**Figure 8**. Doppler spectra generated using wide-beam, plane-wave, and focused-beam transmission. The simulated velocity is approximately 14 cm/s. Wide-beam transmission improved the SNR by approximately 10 dB relative to plane-wave transmission.

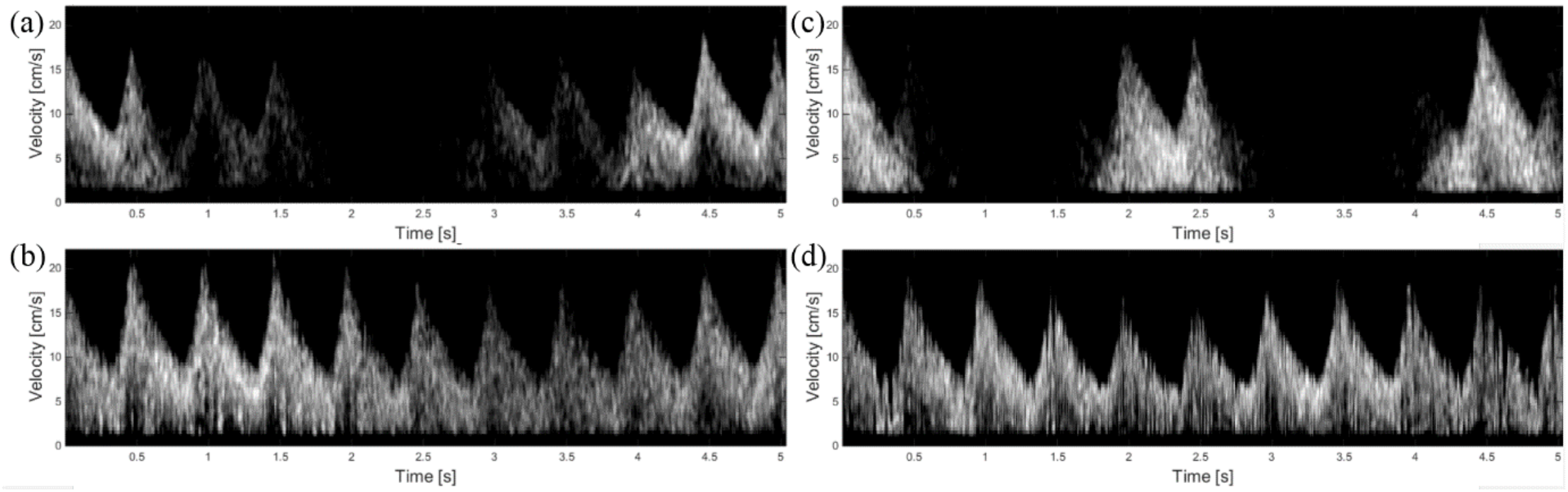

**Figure 9**. Simulated Doppler spectra obtained using (a) conventional processing with axial motion, (b) the proposed wide-beam based method with axial motion, (c) conventional processing with lateral motion, and (d) the proposed wide-beam method with lateral motion.

In the Field II simulation that is described in **Section II. C**, flow targets with global motions were simulated, and the new wide-beam was transmitted by virtual transducer. RF data for 2D SV region were reconstructed by parallel received beamforming [8] to generate multi-beams, which is used for conventional and the proposed 2D SV based spectral Doppler processing and the spectrum is shown in **Figure 9**. The flow targets in **Figure (a) and (b)** were moved with the axial global motion, and those in **Figure (c) and (d)** were moved with the lateral global motion. Spectral Doppler images of **the upper and lower row in Figure (9)** were acquired using conventional processing and the proposed motion tracking Doppler with 2D SV, respectively. Hidden spectra in **Figure 9 (a) and (c)** reappear in **Figure 9 (b) and (d)** since the axial and lateral motions are successfully tracked. **Figure 10 (a) and (b)** represent the outputs of classification problems in neural network, such as scanline and sample shift selections, respectively. **Figure 10** plots corresponding neural network results of **Figure 9**, revealing that tracked positions are located inside the area of vessels.

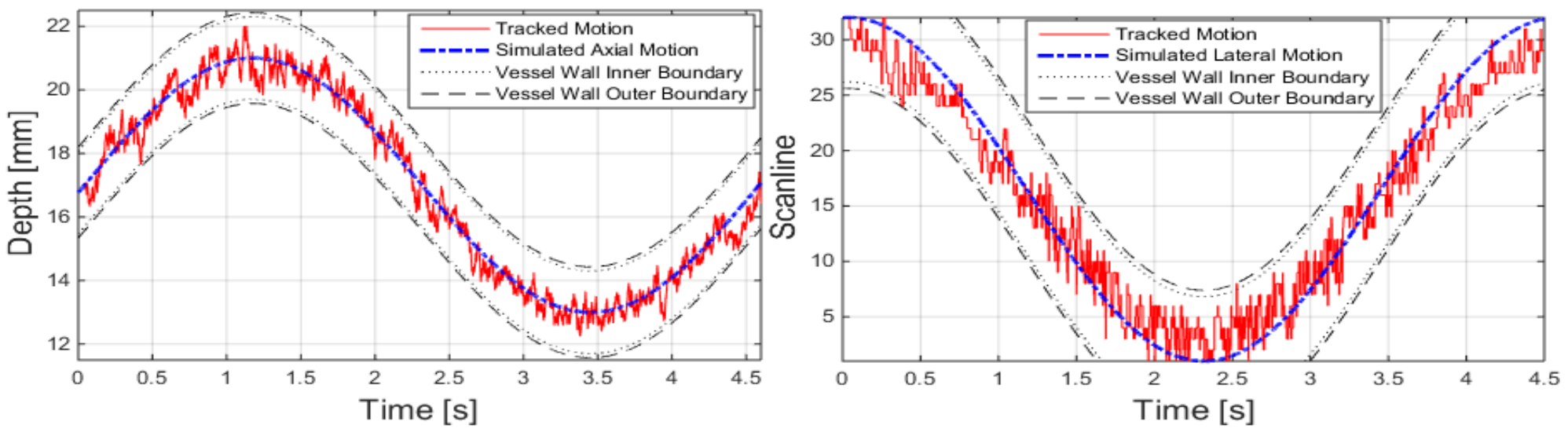

**Figure 10**. Neural network motion tracking results for a vessel radius of 1.5 mm. (a) Sample shift selection during axial processing. (b) Scanline selection during lateral processing.

The wide-beam method enables lateral motion tracking with the neural network, of which the accuracy may depend on the vessel size. **Figure 11** shows the % Correct Classification of lateral motion tracking along with the radius of virtual vessels in Field II. The larger the radius, the higher the accuracy. In case of using the wide-beam, when the vessel is larger than 1 mm of radius, the % Correct Classification is higher than 93 %; furthermore, that is 100% for vessels whose radius is larger than 2 mm. Motion tracking using the wide-beam is more accurate than that of plane wave according to **Figure 11**.

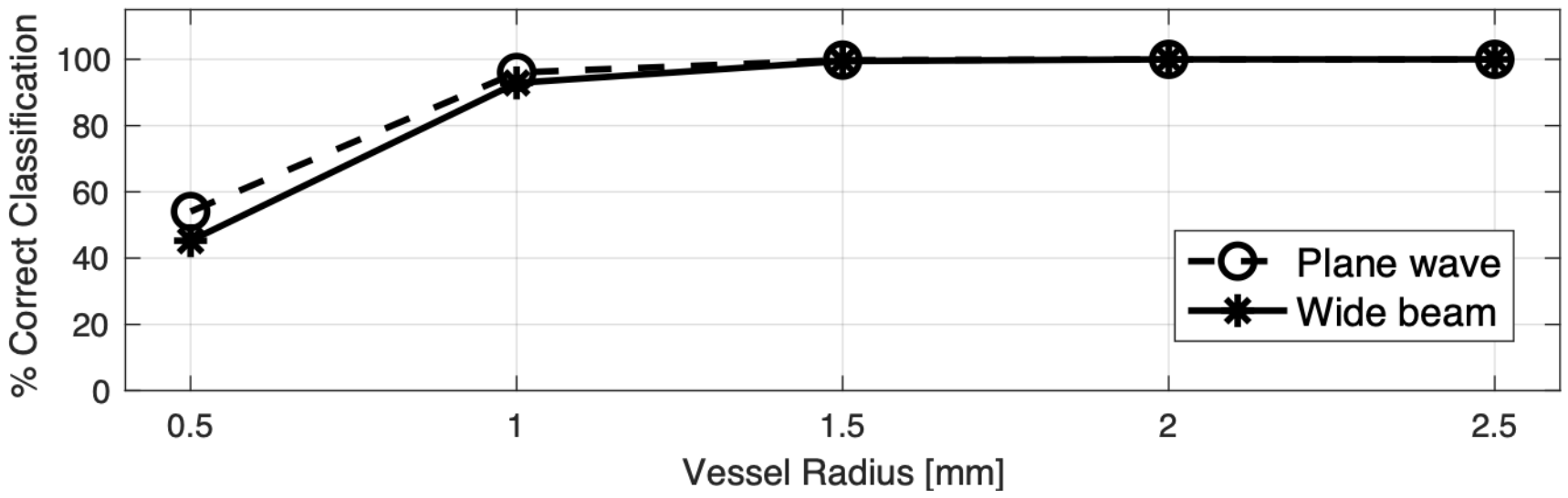

**Figure 11**. % Correction Classification, accuracy of motion tracking as a function of vessel radius for wide-beam and plane-wave acquisition of the 2D SV.

**B. *In vivo* study**

**Figures 12 and 13** compare spectral Doppler images obtained using the conventional and proposed methods for a carotid artery and a hepatic vein scanned with linear and convex transducers, respectively. **Panels (a) and (b)** show the spectra obtained using conventional spectral Doppler processing with a 1D SV and the proposed motion tracking with a 2D SV, respectively. The proposed method reveals previously obscured flow characteristics and increases SNR relative to conventional Doppler processing. **Panel (c)** in each figure represents spectral signal and noise levels at four time points during the scan. The spectra are plotted in dB, with the maximum signal level set to 0 dB. Differences in spectral noise levels between the conventional and proposed methods were assessed over the entire 1.9s scan. The average increases in SNR from the conventional to the proposed method were 9.3 dB for the linear probe and 4.5 dB for the convex probe, as shown in Table 2. The axial and lateral motion tracked by the neural network is plotted in **panels (d) and (e)**, respectively. **Figure 12(d)** shows that the carotid artery moved from a depth of approximately 1.3 cm to 1.8 cm, and **Figure 12(e)** show its lateral tissue motion. Similarly, **Figures 13(d) and 13(e)** show the tracked hepatic vein motion in the depth and scanline directions. The vessel's axial position ranged from 5.8 cm to 6.3 cm, and its lateral position ranged from $4^{th}$ to $8^{th}$ scan lines.

| Application | Carotid artery | Hepatic vein |
|---|---|---|
| Transducer type | Linear | Convex |
| Increase in SNR [dB] | 9.3 | 4.5 |

**Table 2**. Increase in spectral SNR when motion tracking is applied to wide-beam data.

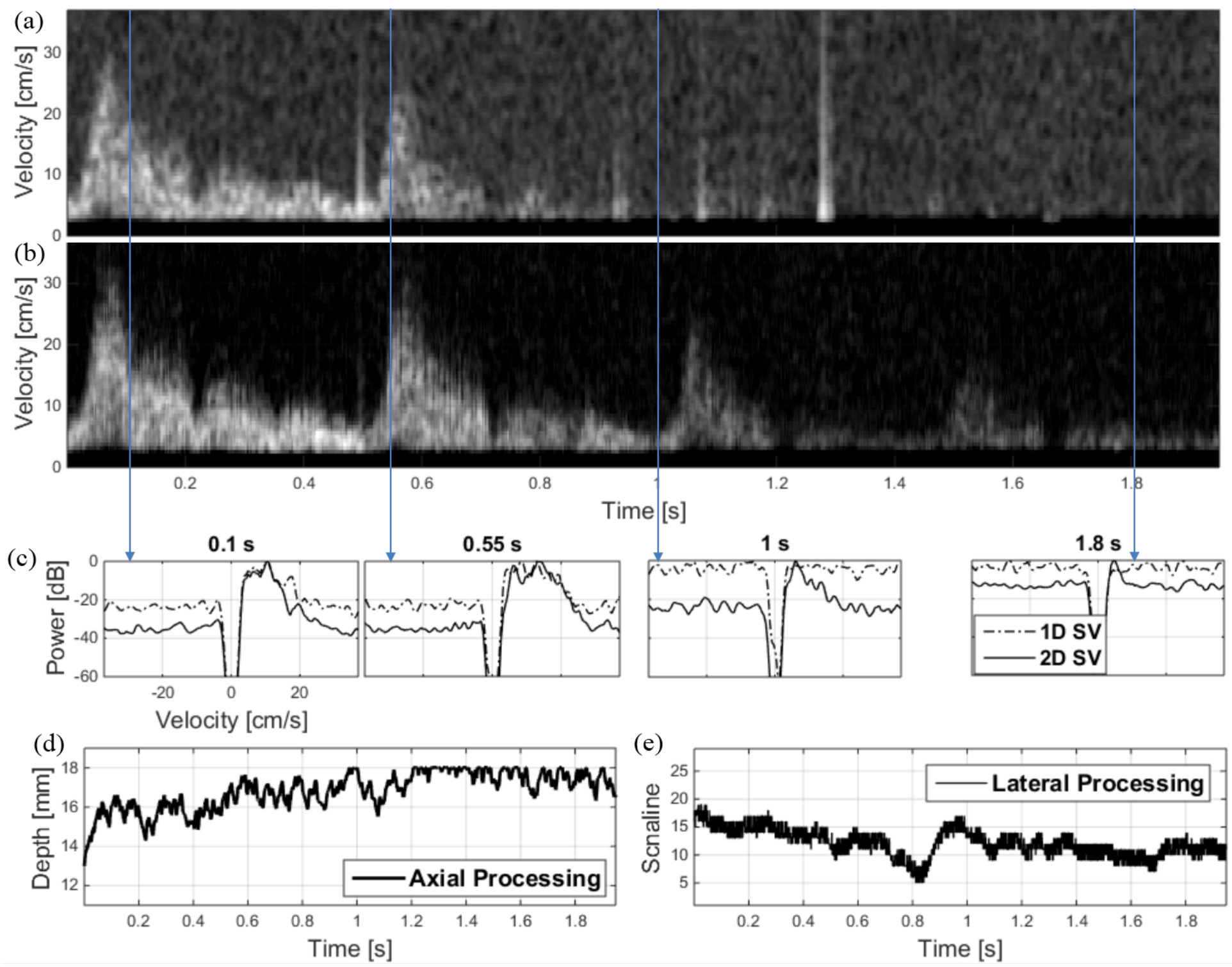

**Figure 12**. Spectral Doppler imaging of the carotid artery. (a) Conventional processing with a fixed 1D SV. (b) Proposed processing with a 2D SV using wide-beam transmission. (c) Comparison of SNR between 1D SV and 2D SV methods; the 2D SV method shows lower relative noise levels. (d) Axial and (e) lateral motion estimated by the neural network using the 2D SV.

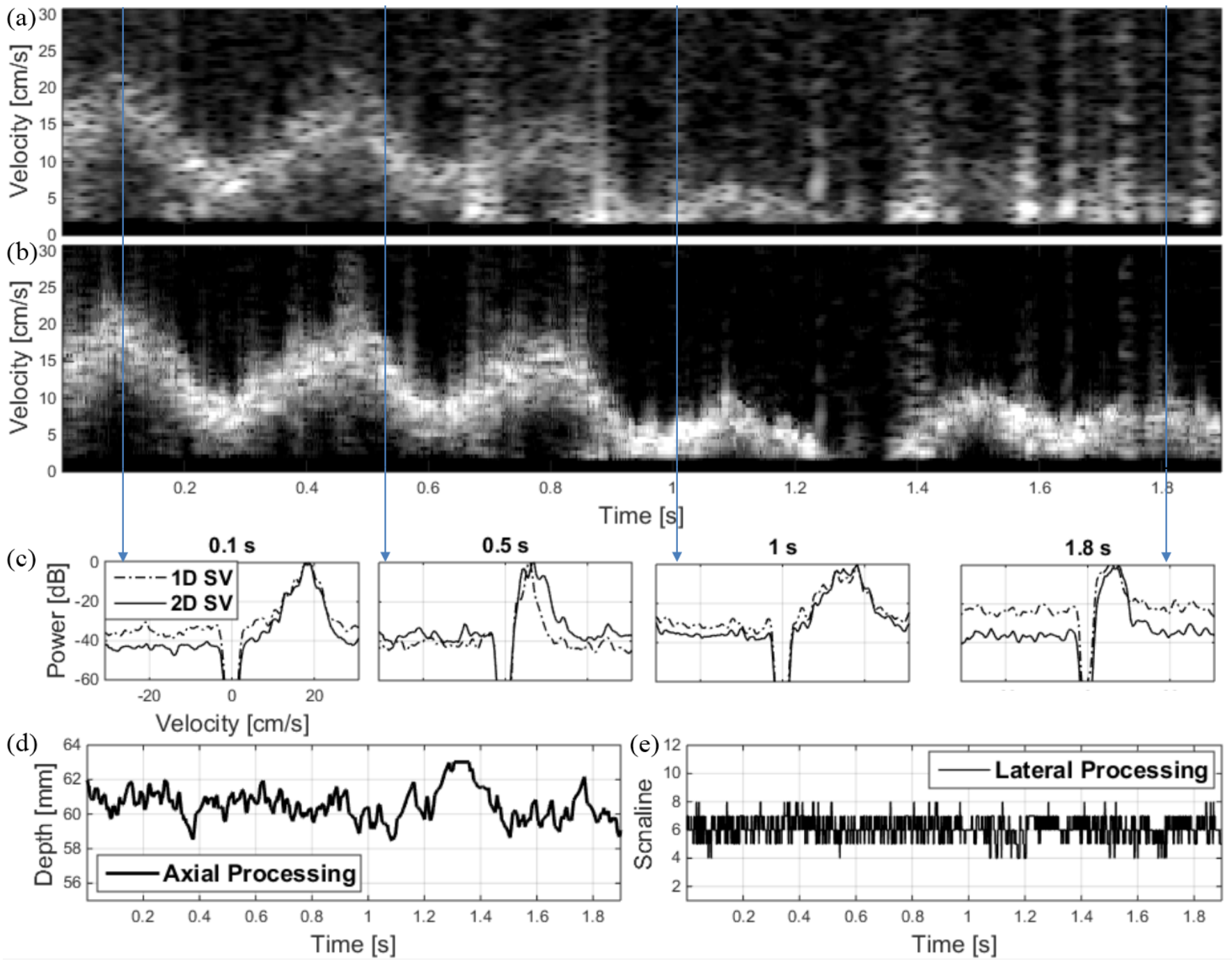

**Figure13**. Spectral Doppler imaging of the hepatic vein. (a) Conventional processing with a fixed 1D SV. (b) Proposed processing with a 2D SV using wide-beam transmission . (c) Comparison of SNR between 1D SV and 2D SV methods; the 2D SV method shows lower relative noise levels. (d) Axial and (e) lateral motion estimated by the neural network using the 2D SV.

## IV. DISCUSSION

This study reveals that the proposed wide-beam method can extend the lateral and axial region of interest (ROI) for velocity estimation in spectral Doppler imaging. We investigated its application with both the linear and convex array transducers and observed improved Doppler spectra and increased spectral SNR with both probe types. Wide-beam transmission was used with the proposed neural network–based motion tracking method to track lateral and axial motion.

The proposed wide-beam method offers several advantages over focused-beam and plane-wave methods for spectral Doppler imaging. First, it provides a 2D SV without reducing the PRF. Focused-beam methods require multiple transmissions to generate 2D data, which reduces the effective PRF. Uncompounded plane-wave transmission have lower transmit energy, resulting in reduced spectral SNR, whereas compounded plane-wave transmission can introduce decorrelation between the compounded data and reduce the effective PRF. Wide-beam transmission requires a single transmit event to generate a 2D SV, avoiding these limitations. Although the lateral ROI of a 2D SV generated by the wide-beam transmission is smaller than obtained with plane-wave transmission, the ROI size is sufficient for spectral Doppler imaging given the human body motions discussed in **Section II** [16, 17]. In addition, the transmit energy is comparable to that of the focused-beam transmission and higher than that of plane-wave transmissions, allowing the new wide-beam method to maintain spectral SNR comparable to that of focused-beam transmission. Finally, the new wide-beam transmission extends the lateral region from which data can be used for spectral Doppler processing, i.e. velocity estimations. This enables tracking of the 2D tissue motion during scanning and can improve spectral accuracy.

The increase in spectral SNR suggests that the proposed wide-beam method can improve PW Doppler imaging. However, a larger SNR increase in **Table 2** does not necessarily indicate better performance because the magnitude of the increase could depend on tissue motion. In other words, greater motion can produce a larger SNR increase, whereas no increase in SNR would be expected in the absence of tissue motion. For instance, **Table 2** shows improvement in spectral Doppler images for both cases.

To evaluate *in vivo* performance, we investigated a carotid artery and a hepatic vein, which differ in vessel size, beam-to-flow angle, and transducer type. Compared with the carotid artery, the hepatic vein has more complex hemodynamics, a smaller diameter, and a wider range of beam-to-flow angles, which require higher-resolution input data and potentially make processing more challenging. However, as shown in **Section III**, the proposed method successfully processed both cases. Because the vessel radii of carotid arteries and hepatic vessels are typically over 1 mm, the *in vivo* results are consistent with the simulation outcomes; that is, the accuracy is over 93% when the vessel radii are greater than 1 mm. In our experiments, the carotid artery exhibited greater motion than the hepatic vein; however, this observation is limited to the scans examined here and should not be generalized to all carotid arteries and hepatic veins, because the amount of motion can depend on probe movement and patient respiration/movement during scanning. Further studies could investigate more dynamic cases, such as motion in the kidney.

Tracking motion within the 2D SV requires an appropriate motion tracking method. We formulated 2D motion tracking in spectral Doppler imaging as a classification problem, in which the position of the adaptive 1D SV is determined based on motion. The neural network selects the sample and scanline that define the 1D SV position, thereby expressing motion tracking as a classification task.

We selected a Hamming network for two reasons. First, it requires no prior training, which is useful in medical imaging, where large training datasets can be difficult to obtain. Whereas many neural networks learn weight matrices during training, the Hamming network constructs its weights directly from candidate patterns derived from the input data. Second, its two-layer architecture is computationally simple compared with DNNs. This simplicity is useful for ultrasound imaging, where real-time processing is important.

As described in Section II.B, the motion tracking inputs included tissue signals, blood-flow signals, and their axial derivatives. Tissue and blood-flow features provide information about speckle and flow motion, whereas derivatives provide additional information about local signal changes. Because ultrasound data are typically sampled more densely in the axial direction than in the lateral direction, we calculated axial rather than lateral derivatives. Axial derivatives may also help distinguish the boundaries of structures that extend laterally across the image. As most tissue structures have horizontal structures rather than vertical structures in ultrasonography, derivatives in the axial direction is, therefore, likely to result in clearer and more detailed performance than that in the lateral direction.

The current method tracks motion within a 2D SV and cannot follow out-of-plane motion in the elevational direction. It could be extended to 3D by using a 2D array transducer to acquire a 3D SV and applying analogous position selection in three dimensions.

## V. CONCLUSION

This study proposed wide-beam transmission method to acquire a 2D SV and a Hamming network based motion tracking method for spectral Doppler imaging. Wide-beam transmission enables acquisition of a 2D SV from a single transmit event while preserving PRF for Doppler processing. Field II simulations and *in vivo* studies demonstrated the feasibility of tracking a moving target, including motion from respiration and transducer probe movement. This study suggests the potential of using Hamming network for real-time 2D motion tracking in spectral Doppler imaging without neural network training or learning.